\documentclass[12pt]{iopart}

\usepackage{iopams}
\usepackage{graphics}
\usepackage{epsfig}
\usepackage[usenames]{color}
\begin{document}

\title[Fr\"ohlich Polarons]{Fr\"ohlich Polarons from 0D to 3D: Concepts and Recent Developments\footnote{The workshop ``Mott's Physics in Nanowires and Quantum Dots'' provides me with an opportunity to recall  stimulating discussions, correspondence and encounters {which} I had with
Sir Nevill, i.~a. in Cambridge, in Antwerpen, in Leuven,  in Roma...
He presented a fine series of lectures : {\it ``Solved and unsolved problems for non-crystalline materials''}
in Antwerpen in July 1975. I enjoyed discussions on small polarons and hopping, on the stability of bipolarons,
on electronic transport in general. Nevill often approached a problem deeper and deeper by a kind of {\it Plato-type} approach with ever more penetrating questions \cite{Mott1973,Mott1990,Devreese1963}. {Some aspects of the present paper are treated in more generality in the chapter by the present author in the book \cite{ALBOOK}. I also refer to the contributions by A.~S.~Alexandrov; S.~Aubry; H.~B\"ottger, V.~V.~Bryksin and T.~Damker; A.~M.~Bratkovsky; V.~Cataudella, G.~De~Filippis and C.~A.~Perroni; H.~Fehske and S.~A.~Trugman; Yu.~A.~Firsov; M.~Hohenadler and W.~von~der~Linden; V.~V.~Kabanov; P.~Kornilovitch; D.~Mihailovic; A.~S.~Mishchenko and N.~Nagaosa; Guo-meng~Zhao; M.~Zoli to Ref. \cite{ALBOOK}. See further the article by A. M. Stoneham et al. \cite{Stoneham}.}}}

\author{J T Devreese}

\address{{Universiteit Antwerpen}, Groenenborgerlaan 171, B-2020 Antwerpen, Belgium
and Technische Universiteit Eindhoven, P. O. Box 513, NL-5600 MB Eindhoven, The Netherlands}
\ead{jozef.devreese@ua.ac.be}
\begin{abstract}
I analyse our present understanding of the Fr\"ohlich polaron with emphasis on the response properties, in particular optical absorption.
%The main topics are: the  Fr\"ohlich  polaron concept; polaron response in 3D; many-polaron systems in 3D and in 2D;
%polarons in 2D and in quasi-2D structures; polarons in quasi-0D structures.
\end{abstract}

%Uncomment for PACS numbers title message
%\pacs{00.00, 20.00, 42.10}
% Keywords required only for MST, PB, PMB, PM, JOA, JOB?
%\vspace{2pc}
%\noindent{\it Keywords}: Article preparation, IOP journals
% Uncomment for Submitted to journal title message
%\submitto{\JPA}
% Comment out if separate title page not required
\maketitle

%\tableofcontents

\section {The  Fr\"ohlich  polaron concept }

As is generally known, the polaron concept was introduced by Landau in 1933 \cite{Landau}. Initial theoretical
\cite{LP48,Pekar,Bogolyubov,BT49,Fr54,LLP53,Feynman} and experimental \cite{B63} works laid {the} fundamental background of polaron physics.
Among the comprehensive review papers and books covering the subject,
I refer to the reviews \cite{KW63,Greenbook,A68,Devreese96,AM96,Calvani}.

Significant extensions and recent developments of the polaron concept have
been realised (see, for example, {\cite{AM1994,Varenna1997,Alexandrov2003,Devreese2005,Varenna2005}}
and references therein). They have been invoked, e. g., to study the properties of conjugated polymers, colossal magnetoresistance perovskites, high-$T_{\mathrm{c}}$ superconductors,
layered MgB$_{\mathrm{2}}$ superconductors, fullerenes, quasi-1D conductors, semiconductor nanostructures.
Polaronic phenomena have been clearly revealed in optical properties of semiconductor quantum wells, superlattices and quantum dots.

A distinction was made between polarons in the continuum approximation where long-range electron-lattice
interaction prevails ({\it Fr\"ohlich}, or {\it large}, polarons)\cite{Pekar,Fr54} and polarons
for which the short-range interaction is essential ({\it Holstein, Holstein-Hubbard} models).
An electron or a hole, trapped by its self-induced atomic (ionic) displacement field in a region of linear
dimension, which is of the order of the lattice constant, is called a {\it small polaron}
\cite{Frohlich1957,Holstein1959}. An excellent survey of the small-polaron physics relevant
to the conduction phenomena in non-crystalline materials and to the metal-insulator transitions has been given
by Mott \cite{Mott1987,Mott1990}.

%\subsection{All-coupling theory: The Feynman path integral. Comparison with the Monte Carlo schemes}

\subsection{Scaling relations for {Fr\"ohlich} polarons in 2D and in 3D}
\label{scaling}

Several scaling relations connect the {Fr\"ohlich} polaron self-energy, the effective
mass, the impedance $Z$ and the polaron mobility $\mu $ in 2D to their counterpart in 3D.
Those relations were obtained by Peeters and Devreese \cite{prb36-4442} at the level of the Feynman model,
keeping the surface phonons for the 2D system, and are listed here:
\begin{eqnarray}
E_{\mathrm{2D}}(\alpha ) &=&\frac{2}{3}E_{\mathrm{3D}}\left( \frac{3\pi }{4}%
\alpha \right),  \label{E} \\
\frac{m_{\mathrm{2D}}^\ast(\alpha)}{m_{\mathrm{2D}}} &=& \frac{m_{\mathrm{3D}%
}^\ast(\frac{3}{4}\alpha)}{m_{\mathrm{3D}}}\ ,  \label{eq:P24-3} \\
Z_{\mathrm{2D}}(\alpha ,\Omega ) &=&Z_{\mathrm{3D}}\left( \frac{3\pi }{4}\alpha
,\Omega \right),
\end{eqnarray}%
where $\Omega $ is the frequency of the external electromagnetic field, and
\begin{equation}
\mu _{\mathrm{2D}}(\alpha )=\mu _{\mathrm{3D}}\left( \frac{3\pi }{4}\alpha
\right).  \label{MU}
\end{equation}

In Eq. (\ref{eq:P24-3}), {$m_{\mathrm{2D}}^\ast$ ($m_{\mathrm{3D}}^\ast$)
and $m_{\mathrm{2D}}$ ($m_{\mathrm{3D}}$) are, respectively, the polaron- and
the electron-band masses in 2D (3D)}. Expressions (\ref{E}) to (\ref{MU})
provide a straightforward link between polaron characteristics in 3D with those in 2D.

The fulfilment of the PD-scaling relation \cite{prb36-4442} is checked here for the
path integral Monte Carlo results \cite{TPC} for the polaron free energy.

\begin{figure}[tbh]
\begin{center}
\includegraphics[height=0.6\textheight]{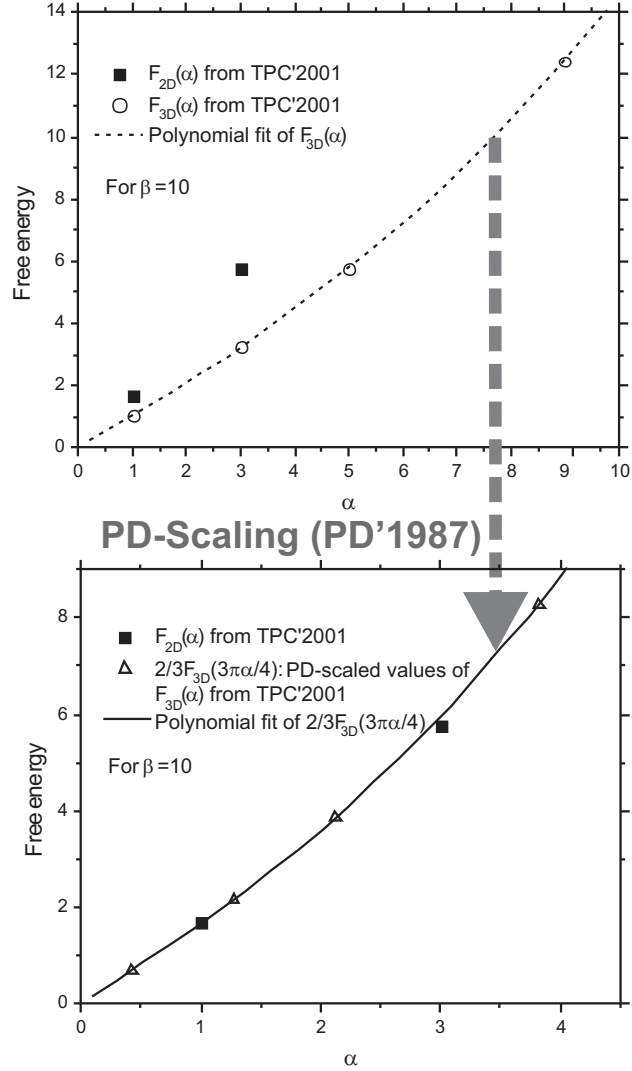}
\end{center}
\caption{\textit{Upper panel: }The values of the polaron free energy in 3D
(squares) and 2D (open circles) obtained by TPC'2001 \cite{TPC} for $\beta
=10$. The data for $F_{3D}\left(  \alpha\right)  $ are interpolated\ using a
polynomial fit to the available four points (dotted line). \textit{Lower
panel:} Demonstration of the PD-scaling cf. PD'1987: the values of the polaron
free energy in 2D obtained by TPC'2001 \cite{TPC} for $\beta=10$ (squares) are
very close to the \textbf{PD-scaled} \cite{prb36-4442}
values of the polaron free energy in 3D from TPC'2001 for $\beta=10$ (open
triangles). The data for $\frac{2}{3}F_{3D}\left(  \frac{3\pi\alpha}%
{4}\right)  $ are interpolated\ using a polynomial fit to the available four
points (solid line).
({Reprinted with permission from \protect\cite{Devreese2005_V}.
\copyright 2006, Societ\`{a} Italiana di Fisica.})}
\label{ScComp}%
\end{figure}

The path integral Monte Carlo results of Ref.\cite{TPC} for the polaron free
energy in 3D and in 2D are given for a few values of temperature and for some
selected values of $\alpha.$ For a check of the scaling relation, the values
of the polaron free energy at $\beta\equiv\hbar\omega_{\mathrm{LO}}/k_{B}T=10$ are taken from Ref. \cite{TPC} in 3D
(Table I, for 4 values of $\alpha$) and in 2D (Table II, for 2 values of
$\alpha$) and plotted in Fig. \ref{ScComp}, upper panel, with squares and open
circles, correspondingly. (Here $\omega_{\mathrm{LO}}$ is the frequency of the long-wavelength
longitudinal optical phonons, $T$ is the temperature).

In Fig. \ref{ScComp}, lower panel, the available data for the free energy from
Ref \cite{TPC} are plotted in the following form \textit{inspired by the
l.h.s. and the r.h.s parts of Eq. (\ref{E})}: $F_{2D}\left(  \alpha\right)  $
(squares) and $\frac{2}{3}F_{3D}\left(  \frac{3\pi\alpha}{4}\right)  $(open
triangles). As follows from the figure, t\textit{he path integral Monte Carlo
results for the polaron free energy in 2D and 3D very closely follow the
PD-scaling relation of the form given by Eq. (\ref{E}):}%
\begin{equation}
F_{2D}\left(  \alpha\right)  \equiv\frac{2}{3}F_{3D}\left(  \frac{3\pi\alpha
}{4}\right)  . \label{F}%
\end{equation}

\section{Bipolarons}

When two electrons (or two holes) interact with each other
simultaneously through the Coulomb force and via the
electron-phonon-electron interaction, either two independent polarons can
occur or a bound state of two polarons --- the {\it bipolaron} --- can arise
\cite{Vinetskii1961,HT1985,Adamowski1989,Bassani1991,Verbist1990,Verbist1991}.
Whether bipolarons originate or not, depends on the competition between
the repulsive forces (direct Coulomb interaction
$U({\bf r}) = \frac{e^2}{\varepsilon_\infty |{\bf r}|} \equiv\frac{U}{|{\bf r}|}$,) and the attractive
forces (mediated through the electron-phonon interaction).
In the discussion of bipolarons often the  ratio
\begin{equation}
\eta = \frac{\varepsilon_\infty}{\varepsilon_0}
\end{equation}
of {the} electronic and static dielectric constant is used ($0\le\eta\le 1$).
It {turns} out that bipolaron formation is favoured by smaller
$\eta$.

The Fr\"ohlich bipolaron was analyzed \cite{Verbist1990,Verbist1991} using the Feynman path
integral formalism. Quite analogously to the above discussed relations (\ref{E}) to (\ref{MU}),
a scaling relation was derived between the free energies $F$ in two dimensions
$F_{2D}(\alpha, U, \beta)$ and in three dimensions $F_{3D}(\alpha, U, \beta)$:
\begin{equation}
F_{2D}(\alpha, U, \beta) =
\frac{2}{3} F_{3D}(\frac{3\pi}{4}\alpha, \frac{3\pi}{4}U, \beta).
\end{equation}
Physically {this} scaling relation implies that bipolaron formation will be
facilitated in 2D as compared to 3D.

\begin{figure}[t]
\begin{center}
\includegraphics[width=0.5\textwidth]{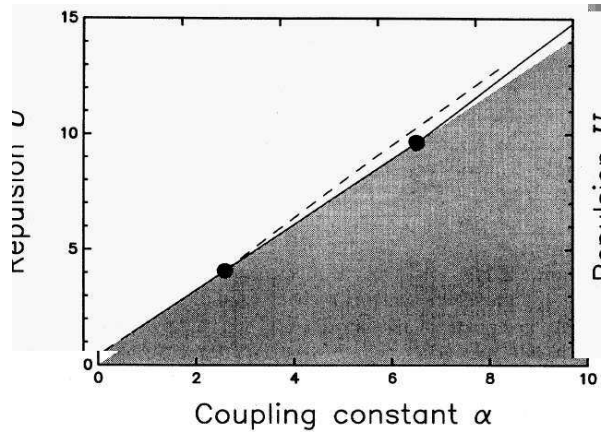}
\end{center}
\caption{The phase diagram for bipolaron formation in two (dashed curve) and
three dimensions (solid curve) is presented. Bipolarons are formed below the
{dashed/solid} curves. The nonphysical part $U \le \sqrt{2}\alpha$ of the $(\alpha,U)$
space is shaded.
({Reprinted with permission from Ref. \protect\cite{Verbist1991}. \copyright 1991 by the American Physical Society}.)}
\label{Bipolaron}%
\end{figure}
A ``phase-diagram'' for the polaron---bipolaron system was
introduced in Refs.\cite{Verbist1990,Verbist1991}.
It is based on {a} generalized trial action. This
phase diagram is shown in Fig.\,\ref{Bipolaron}.  For the 3D-case, a Fr\"ohlich coupling
constant as high as 6.8 is needed to allow for bipolaron formation. No
definite experimental evidence has been provided for the existence of
materials with such high Fr\"ohlich coupling constant.
Materials with  sufficiently  large $\alpha$  for Fr\"ohlich bipolaron
formation in 3D might  exist, but careful analysis
(involving e.~g. the study of cyclotron resonance), is needed in order to confirm this.
{Undoubtedly,} some modifications to the Fr\"ohlich Hamiltonian are also necessary to
describe such high coupling because of the more localized character of
the carriers in this case which makes the continuum approximation less valid.
The confinement of the bipolaron in 2D facilitates bipolaron
formation at smaller $\alpha$. From Fig.\,\ref{Bipolaron} it is seen that bipolarons
in 2D can be stable for $\alpha\ge 2.9$, a domain of coupling constants
which is  definitely  realized in several solids.

At present no consensus exists concerning the microscopic mechanism governing the creation
of Cooper pairs in high-T$_c$ superconductors. Nevertheless, there is definitely evidence
for the existence of polarons and bipolarons at least in the normal phase of high-T$_c$ superconductors
as manifested, e. g., in their optical absorption spectra (see \cite{Devreese2005} and references therein).

\section{Polaron response in 3D}

%\subsection{Polaron mobility}

\subsection{Optical absorption of a polaron at strong coupling}

The problem of the structure of the large polaron excitation spectrum
constituted a central question in the early stages of the development of
polaron theory. The exactly solvable polaron model of Ref. \cite{DThesis}
was used to demonstrate the existence of the so-called \textquotedblleft
relaxed excited states\textquotedblright of large polarons \cite{DE64}. In
Ref. \cite{R65}, and after earlier intuitive analysis, this problem was
studied using the classical equations of motion and Poisson-brackets. The
insight gained as a result of those investigations concerning the structure
of the excited polaron states, was subsequently used to develop a theory of
the optical absorption spectra of polarons. The first work was limited to
the strong coupling limit \cite{KED69}. Ref. \cite{KED69} is the first work
that reveals the impact of the internal degrees of freedom of polarons on
their optical properties.

The optical absorption of light by free Fr\"ohlich polarons was treated in
Ref.~\cite{KED69} using the polaron states obtained within the adiabatic
strong-coupling approximation. It was argued in Ref.~\cite{KED69}, that for
sufficiently large $\alpha$ ($\alpha>3$), the (first) relaxed excited state
(RES) of a polaron is a relatively stable state, which can participate in
optical absorption transitions. This idea was necessary to understand the
polaron optical absorption spectrum in the strong-coupling regime. The
following scenario of a transition, which leads to a \textit{%
\textquotedblleft zero-phonon\textquotedblright\ peak} in the absorption by
a strong-coupling polaron, then {was} suggested. If the frequency of the incoming photon is equal to $\Omega_{\mathrm{RES}}=0.065\alpha^{2}\omega_{\mathrm{LO}},$ the electron jumps from the ground state (which, at {sufficiently} large
coupling, is well-characterized by "$s$"-symmetry for the electron) to an
excited state ("$2p$"), while the lattice polarization in the final state is
adapted to the "$2p$" electronic state of the polaron. In Ref. \cite{KED69},
considering the decay of the RES with emission of one real phonon, it is
argued, that the \textquotedblleft zero-phonon\textquotedblright\ peak can
be described using the Wigner-Weisskopf formula valid when the linewidth of
that peak is much smaller than $\hbar\omega_{\mathrm{LO}}.$

For photon energies larger than $\Omega_{\mathrm{RES}}+\omega_{\mathrm{LO}},$
a transition of the polaron towards the first scattering state, belonging to
the RES, becomes possible. The final state of the optical absorption process
then consists of a polaron in its lowest RES plus a free phonon. A
\textquotedblleft one-phonon sideband\textquotedblright\ then appears in the
polaron absorption spectrum. This process is called \textit{one-phonon
sideband absorption}. The one-, two-, ... $K$-, ... phonon sidebands of the
zero-phonon peak give rise to a broad structure in the absorption spectrum.
It turns out that the \textit{first moment} of the phonon sidebands
corresponds to the Franck-Condon (FC) frequency $\Omega_{\mathrm{FC}%
}=0.141\alpha^{2}\omega_{\mathrm{LO}}.$ To summarize, the polaron optical
absorption spectrum at {sufficiently} strong coupling is characterized by the following
features (at $T=0$):

\begin{enumerate}
\item[a)] An absorption peak (\textquotedblleft zero-phonon
line\textquotedblright) appears, which corresponds to a transition from the
ground state to the first RES at $\Omega_{\mathrm{RES}}$.

\item[b)] For $\Omega>\Omega_{\mathrm{RES}}+\omega_{\mathrm{LO}}$, a phonon
sideband structure arises. This sideband structure peaks around $\Omega _{%
\mathrm{FC}}$.
\end{enumerate}

{Note that, at $T=0$, the polaron optical absorption exhibits a zero-frequency {feature} $\propto \delta(\Omega).$}

\subsection{Optical absorption of a polaron at arbitrary coupling}

Although the optical conductivity (OC) of the Fr\"ohlich polaron model attracted attention for years \cite{Devreese96}, there exists no {exact} analytic {expression for it at} all coupling. The most
successful approach is that based on the Feynman path integral
technique {as applied in} \cite{DSG72} (DSG) {and \cite{Devreese72}}, where the OC is calculated starting from the Feynman variational model (FVM) \cite{Feynman} for the
polaron and using {Feynman's} path integral response formalism \cite{FHIP}.
Subsequently the path integral approach was rewritten in terms of
the memory function formalism (MFF) \cite{PD1983}. These
approaches are completely successful at small electron-phonon coupling and are
able to identify {key-}excitations at intermediate and
strong electron-phonon coupling. A comparison of the DSG results with  the OC spectra {derived with} recently developed approximation-free numerical Diagrammatic Quantum Monte Carlo (DQMC)
\cite{Mishchenko2003} and approximate analytical approaches has been recently performed in Ref. \cite{PRL2006}.

The polaron absorption coefficient $\Gamma(\Omega)$ of light with frequency $%
\Omega$ at arbitrary coupling was first derived in Ref.\,\cite{DSG72} (see
also \cite{PD1983}). It was represented in the form
\begin{equation}
\Gamma(\Omega)=-\frac{1}{\bar{n}\epsilon_{0}c}\frac{e^{2}}{m_{b}}\frac {\mathrm{Im}%
\Sigma(\Omega)}{\left[ \Omega-\mathrm{Re}\Sigma(\Omega)\right] ^{2}+\left[
\mathrm{Im}\Sigma(\Omega)\right] ^{2}}\ ,  \label{eq:P24-2}
\end{equation}
{where $\epsilon_{0}$ is the dielectric permittivity of the vacuum and ${\bar n}$ is the refractive index of the medium.}
This expression was the starting point for a derivation of the
theoretical optical absorption spectrum of a single Fr\"ohlich polaron at
\textit{all electron-phonon coupling strengths} in Ref.\thinspace\cite{DSG72}%
. $\Sigma(\Omega)$ is the so-called \textquotedblleft memory
function\textquotedblright, which contains the dynamics of the polaron and
depends on $\Omega$, $\alpha$ and temperature. The key contribution of the
work in \cite{DSG72} was to introduce $\Gamma(\Omega)$ in the form (\ref%
{eq:P24-2}) and to calculate $\mathrm{Re}\Sigma(\Omega)$, which is
essentially a (technically not trivial) Kramers--Kronig transform of the
function $\mathrm{Im}\Sigma(\Omega)$. The function $\mathrm{Im}%
\Sigma(\Omega)$ had been formally derived for the Feynman polaron {in Ref.} \cite{FHIP} {where} the polaron mobility $\mu$ {was found} from the impedance function, i.~e.
the static limit
\[
\mu^{-1}=\lim\limits_{\Omega \to 0} \left(\frac{\mathrm{Im}\Sigma(\Omega)}{%
\Omega}\right).
\]

The nature of the polaron excitations was {analysed} through this
polaron optical absorption obtained in \cite{DSG72,PD1983}. It was
{confirmed} in \cite{DSG72} that the Franck-Condon states for Fr\"ohlich polarons {at $3 \lesssim \alpha \lesssim 7$} are nothing else but a superposition of phonon sidebands {(of the RES)}. It was {found} in \cite{DSG72} that a relatively large value of the electron-phonon coupling strength ($\alpha > 3$) is needed to stabilise the relaxed excited state of the polaron. At weaker coupling {($0 < \alpha \lesssim 3$)} only \textquotedblleft scattering states\textquotedblright of the polaron, {in its ground state,} play a role in the optical absorption \cite{DSG72,DDG71}.

In the weak coupling limit, the {optical absorption} spectrum (\ref{eq:P24-2}%
) of the polaron is determined by the absorption of radiation energy, which
is reemitted in the form of LO phonons. For $\alpha\gtrsim {3}$, the polaron
can {undergo} transitions toward a relatively stable RES (see Fig.~\ref%
{fig_3}). The RES peak in the optical absorption spectrum also has a phonon
sideband-structure, whose average transition frequency can be related to a
FC-type transition. Furthermore, at zero temperature, the optical absorption
spectrum of a polaron exhibits also a zero-frequency \textquotedblleft
central peak\textquotedblright\ [$\propto\delta(\Omega)$]. For non-zero
temperature, this \textquotedblleft central peak\textquotedblright\ smears
out and gives rise to an \textquotedblleft anomalous\textquotedblright\
Drude-type low-frequency component of the optical absorption spectrum.

\begin{figure}[ht]
\begin{center}

\includegraphics[width=0.5\textwidth]{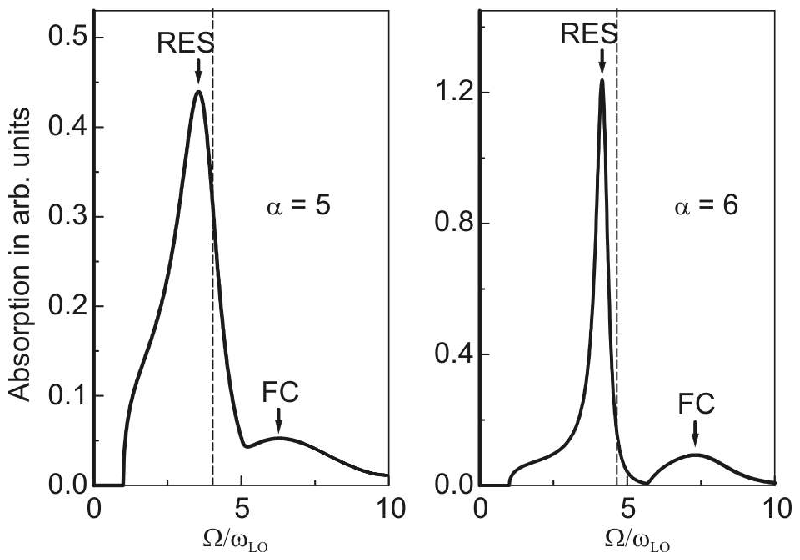}
\end{center}
\caption{Optical absorption spectrum of a Fr\"ohlich polaron for $\protect%
\alpha=5$ and $\protect\alpha = 6$. The RES peak is {relatively} intense compared with the FC peak {at these coupling strengths}. The frequency $\Omega/\protect\omega_{\mathrm{LO}}=v$ is
indicated by the dashed lines. The $\protect\delta$-like central peaks (at $%
\Omega=0$) are schematically shown by vertical lines.
({Reprinted with permission from Ref. \protect\cite{DSG72}. \copyright 1972 by the American Physical Society}.)}
\label{fig_3}
\end{figure}

In Fig. \ref{fig_3} (from Ref. \cite{DSG72}), the main peak of
the polaron optical absorption for $\alpha =$ 5 at $\Omega =3.51\omega _{%
\mathrm{LO}}$ is interpreted as due to transitions to a RES. The
\textquotedblleft shoulder\textquotedblright\ at the low-frequency side of
the main peak is attributed to {mainly} one-phonon transitions to
polaron-\textquotedblleft scattering states\textquotedblright . The broad
structure centered at about $\Omega =6.3\omega _{\mathrm{LO}}$ is
interpreted as a FC band. As seen from Fig. \ref{fig_3}, when increasing the
electron-phonon coupling constant to $\alpha $=6, the RES peak at $\Omega
=4.3\omega _{\mathrm{LO}}$ stabilizes. It is in Ref. \cite{DSG72} that an
all-coupling optical absorption spectrum of a Fr\"{o}hlich polaron, together
with the role of RES-states, FC-states and scattering states, was first
presented. {At larger $\alpha$ (say, for $\alpha > 6$), the linewidths of the absorption peaks obtained in \cite{DSG72} are much too narrow. This shortcoming was noted in \cite{DSG72}.}

Recent numerical calculations of the optical conductivity for
the Fr\"{o}hlich polaron, performed within a Diagrammatic Quantum Monte
Carlo (DQMC) method \cite{Mishchenko2003},
confirm the essential analytical results derived in Ref. \cite{DSG72} for $%
\alpha\lesssim 3.$ In the intermediate coupling regime $3<\alpha<6,$ the
low-energy behaviour and the position of the RES-peak in the optical
conductivity spectrum of Ref. \cite{Mishchenko2003} follow closely the
prediction of Ref. \cite{DSG72}. There are some minor
differences between the two {treatments} in the intermediate coupling regime {$3 \lesssim \alpha \lesssim 6$}:
in Ref. \cite{Mishchenko2003}, the dominant (\textquotedblleft
RES\textquotedblright) peak is less intense in the Monte-Carlo numerical
simulations and the second (\textquotedblleft FC\textquotedblright) peak
develops less prominently. {More significant} differences
between the two {treatments appear} in the strong coupling regime: in Ref. \cite{Mishchenko2003}, the dominant peak broadens and the second peak does not develop, giving instead rise to a flat shoulder in the optical conductivity spectrum at $\alpha=6{.5}.$ These qualitative differences from the optical absorption spectrum of Ref. \cite{DSG72} can be attributed to optical processes with participation of two \cite{Goovaerts73} or more phonons. The above differences can arise also due to the fact that, starting from the Feynman polaron model, one-phonon processes are assigned {relatively} more oscillator strength and the RES of \cite{DSG72} therefore tends to be more stable as compared to the Monte-Carlo result.

\begin{figure}[ht]
\begin{center}
\includegraphics[width=0.5\textwidth]{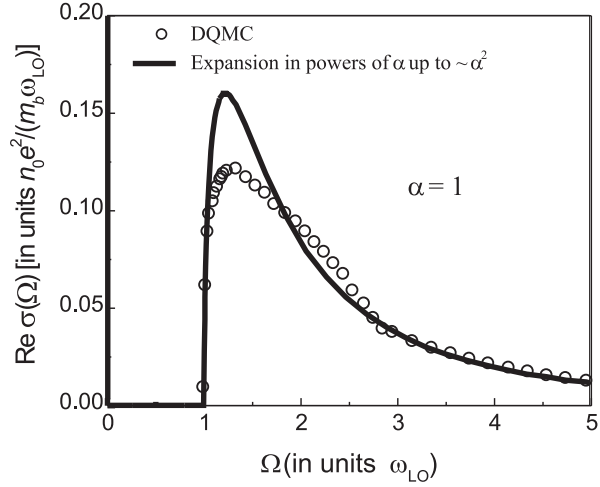}
\end{center}
\caption{One-polaron optical conductivity $\mathrm{Re}\sigma\left(
\Omega\right)  $ for $\alpha=1$ calculated within the DQMC approach
\cite{Mishchenko2003} (open circles) and derived using the expansion in powers
of $\alpha$ up to $\alpha^{2}$ \cite{Huybrechts1973} (solid curve).}%
\label{fig_4a}%
\end{figure}
In Fig. \ref{fig_4a}, {DQMC} optical conductivity spectrum of one polaron for $\alpha=1$ {is compared} with that obtained in Ref. \cite{Huybrechts1973} {where a} canonical-transformation formalism {is used} taking into account correlations in processes involving two LO phonons.

%%%%%%%%%%%%%%%%%%%%%%%%%%%%
\begin{figure}[hb]
\centering \includegraphics[width=1.0\textwidth]{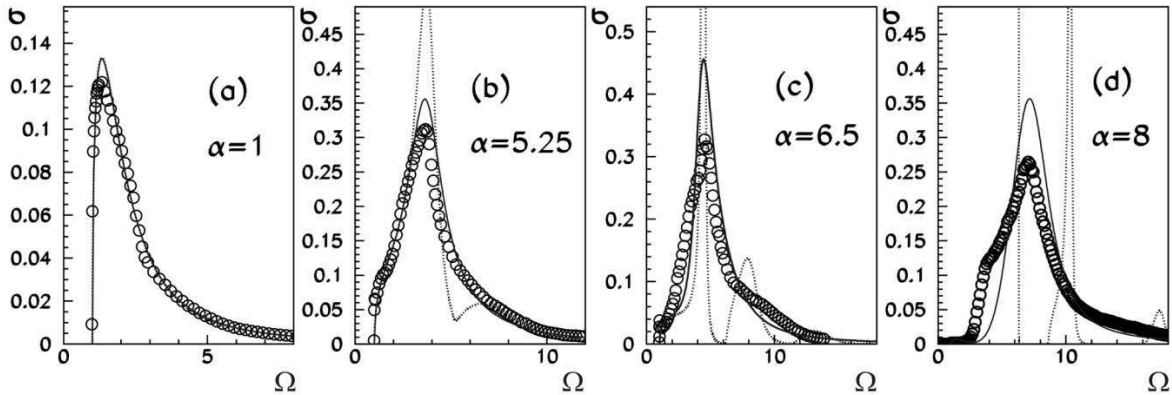}
\caption{{Comparison of the optical conductivity calculated
with the DQMC method (circles), {the} extended MFF (solid line) and DSG \protect\cite{DSG72,Devreese72} (dotted line), for four different values of $\protect\alpha $.
({Reprinted with permission from Ref. \protect\cite{PRL2006}. \copyright 2006 by the American Physical Society}.)}}
\label{figa_graph}
\end{figure}
\begin{figure}[ht]
\centering \includegraphics[width=1.0\textwidth]{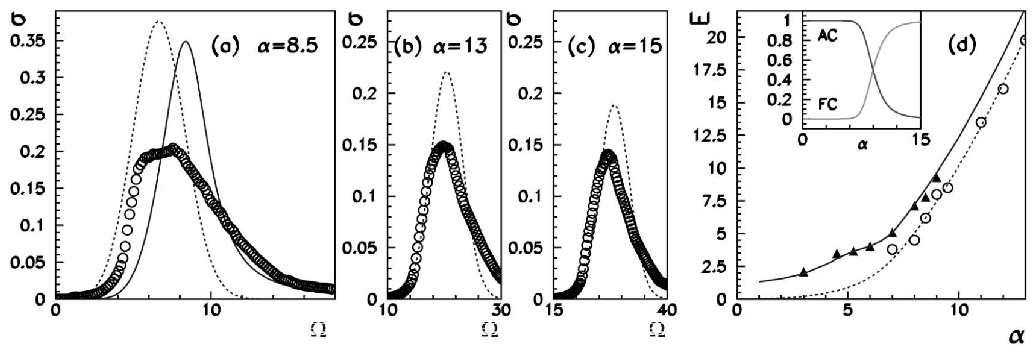}
\caption{{a), b) and c) Comparison of the optical conductivity
calculated with the DQMC method (circles), the extended MFF (solid line)
and SCE (dashed line) for three different values of $\protect\alpha $. d)
The energy of the lower- and higher-frequency features obtained by DQMC
(circles and triangles, respectively) compared (i) with the FC
transition energy  calculated from the SCE (dashed line) and (ii)
with the energy of the peak obtained from the extended MFF (solid line). In
the inset the weights of  Franck-Condon and adiabatically connected
transitions are shown as a function of $\protect\alpha $. We have used for $%
\protect\eta $ the value $1.3$.
({Reprinted with permission from Ref. \protect\cite{PRL2006}. \copyright 2006 by the American Physical Society}.)}}
\label{figb_graph}
\end{figure}
%%%%%%%%%%%%%%%%%%%%%%%%%%%%

The optical conductivity spectra obtained with the
DQMC method {\cite{Mishchenko2003,PRL2006,MishchenkoL}}, extended memory-function formalism (MFF) {\cite{PRL2006,CataudellaL}}, strong-coupling expansion (SCE) {\cite{PRL2006,CataudellaL,DK2006}} and DSG {\cite{DSG72,Devreese72}} for different values of $\alpha $ is shown in Figs. \ref{figa_graph}, \ref{figb_graph}.
The {key} results of the comparison between them are the following. First, as expected, in the weak coupling regime, both the extended MFF with phonon broadening {\cite{PRL2006,CataudellaL}} and DSG {\cite{DSG72,Devreese72}} are in {excellent} agreement with the DQMC data \cite{Mishchenko2003}, showing significant improvement with respect
to the weak coupling perturbation approach \cite{GLF62} which provides a good
description of the OC spectra only for {$\alpha \ll 1$}. For $4\le \alpha \le 8$, where DSG underestimates the peak width, the damping, introduced in the extended MFF approach, {is significant}. Comparing the peak and shoulder energies, obtained by DQMC, with the peak energies, given by {the extended} MFF, and the FC transition energies from the SCE, it
is concluded \cite{PRL2006} that as $\alpha$ increases from $6$ to $10$ the
spectral weights rapidly switch from the dynamic regime, where the lattice
follows the electron motion, to the adiabatic regime dominated by FC
transitions. In the coupling {range} $6 <\alpha <10 $, adiabatic FC and nonadiabatic dynamical excitations coexist. {As $\alpha > 10$ the polaron OC spectrum consists of a broad FC peak, composed of LO phonon sidebands, as proposed in \cite{KED69}, where only 1 LO phonon sideband was calculated.}

\begin{figure}[b]
\begin{center}
\includegraphics[width=0.5\textwidth]{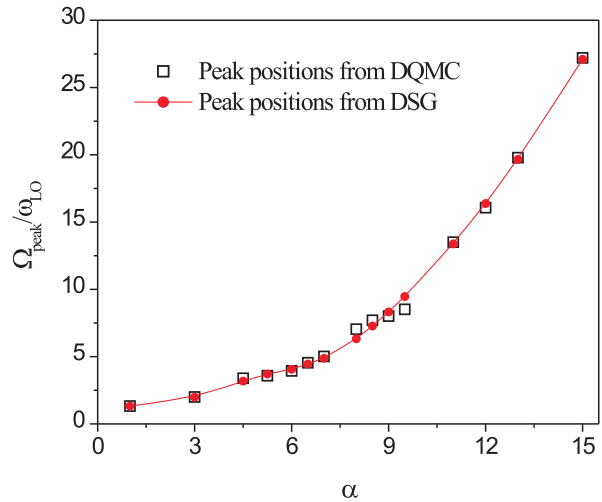}
\end{center}
\caption{Main-peak positions in the DQMC optical conductivity spectra of continuum polarons \cite{PRL2006}
compared to the analytical DSG approach \cite{DSG72}.
(From Ref.\thinspace\cite{DK2006}.)}%
\label{fig_positions}%
\end{figure}

An instructive comparison between the {frequency} of the main peak in the optical absorption spectrum of
continuum polarons obtained within the DSG and DQMC approaches {for various $\alpha$} has been {made} in Ref. \cite{DK2006}.
In Fig. \ref{fig_positions} the energy of the main peak in the OC spectra calculated within the DSG approach {\cite{DSG72,Devreese72}} is plotted together with that given by DQMC {\cite{Mishchenko2003,PRL2006}}.
As seen from the figure, the main-peak positions, obtained within DSG, are in good agreement with the results of DQMC
for all considered values of $\alpha$, {up to $\alpha = 15$}.
The difference between the DSG and DQMC results is {slightly} larger at $\alpha$ = 8 and for $\alpha$ = 9.5, but even for those values of the coupling constant the agreement is {close}.

In summary, the {rigorous} numerical results, obtained using DQMC {\cite{Mishchenko2003,MishchenkoL}},  -- modulo the linewidths for $\alpha > 6$ -- and the recent analytical approximations ({the extended} MFF, SCE) {\cite{PRL2006,CataudellaL,DK2006}} confirm the essence of the mechanism for the optical absorption of Fr\"ohlich polarons, which we proposed in {1969 -} 1971 {\cite{KED69,DSG72,Devreese72}}.

\subsection{Polaron cyclotron resonance}

The RES-like resonances in $\Gamma (\Omega )$, Eq. (\ref{eq:P24-2}),
due to the zero's of $\Omega -\mathrm{Re}\Sigma (\Omega )$, can {\it effectively} be displaced to smaller polaron coupling by applying an external magnetic field $B$, in which case the {condition} for a
resonance becomes $\Omega
-\omega_{c}-\mathrm{Re}\Sigma (\Omega )=0$ ($\omega_{c}=eB/m_{b}c$ is the
cyclotron frequency). Resonances in the magnetoabsorption governed by this
{condition} have been {clearly} observed and analysed.

The results {for} the polaron optical absorption \cite{DSG72,Devreese72} paved
the way for an all-coupling path-integral based theory of {the}
magneto-optical absorption {of} polarons (see Ref. \cite{PD86}).
This work was motivated {i.~a.} by the insight that magnetic
fields can stabilise relaxed excited polaron states, so that information
on the nature of relaxed excited states might be gained from the cyclotron
resonance of polarons.

Some of the subsequent developments in the field of polaron cyclotron
resonance are discussed below. %%%
%\subsection{Cyclotron resonance of polarons in silver halides}

Evidence for the polaron character of charge carriers in AgBr and AgCl
was obtained through high-precision cyclotron resonance experiments in
magnetic fields up to 16 T (see \cite{Hodby1987}). The quantitative
interpretation of the cyclotron resonance experiment in AgBr and AgCl \cite{Hodby1987}
by the theory of Ref. \cite{PD86} provided one of the most convincing and clearest demonstrations of {Fr\"ohlich} polaron features in solids.

The analysis in Ref.\,\cite{Hodby1987} leads to the following
polaron coupling constants: $\alpha=1.53$ for AgBr and $\alpha=1.84$ for
AgCl. The corresponding polaron masses are: $m^*=0.2818m_e$ for AgBr and $%
m^*=0.3988m_e$ for AgCl {($m_e$ is the mass of the free electron)}. For most materials with relatively large Fr\"ohlich
coupling constant {the band mass} $m_b$ is not known. The study in Ref.\,\cite{Hodby1987} is
an example of the detailed analysis of the cyclotron resonance data that is
necessary to obtain accurate polaron data like $\alpha$ and $m_b$ for a
given material.

%\subsection{Cyclotron resonance of polarons in CdTe}

Early infrared-transmission studies of hydrogen-like shallow-donor-impurity
states in $n$-CdTe were reported in Ref.\,\cite{Cohn1972}. By studying the
Zeeman splitting of the ($1s\rightarrow 2p,m=\pm 1$) transition in the
Faraday configuration at magnetic fields up to $\sim $ {16 T}, a
quantitative determination of polaron shifts of the energy levels of a bound
electron was performed. The experimental data were shown to be in fair
agreement with the weak-coupling theory of the polaron Zeeman effect. In
this comparison, however, the value $\alpha =0.4$ had to be used instead of $%
\alpha =0.286$, which would follow from the definition of the Fr\"ohlich
coupling constant. Similarly, the value $\alpha \sim 0.4$ was suggested (see
Ref.~\cite{Larsen72}) for the explanation of the measured variation of the
cyclotron mass with magnetic field in CdTe. This discrepancy gave rise to
some discussion in the literature (see, e. g., Refs.\,\cite%
{Pfeffer88,Miura97} and references therein).

In Ref.\,\cite{Grynberg1996}, far-infrared photoconductivity techniques were
applied to study the energy spectrum of shallow In donors in CdTe layers and
experimental data were obtained on the magnetopolaron effect. An overall good
agreement {was} found between experiment
and a theoretical approach, where the electron-phonon interaction is treated
within a second-order improved Wigner-Brillouin perturbation theory and a
variational calculation is performed for the lowest-lying donor states ($1s,
2p^{\pm}, 2s, 2p_z, 3d^{-2}, 4f^{-3 }$). This agreement {was} obtained with the coupling constant $\alpha=0.286$.

\begin{figure}[htbp]
\begin{center}
\includegraphics[width=0.8\textwidth]{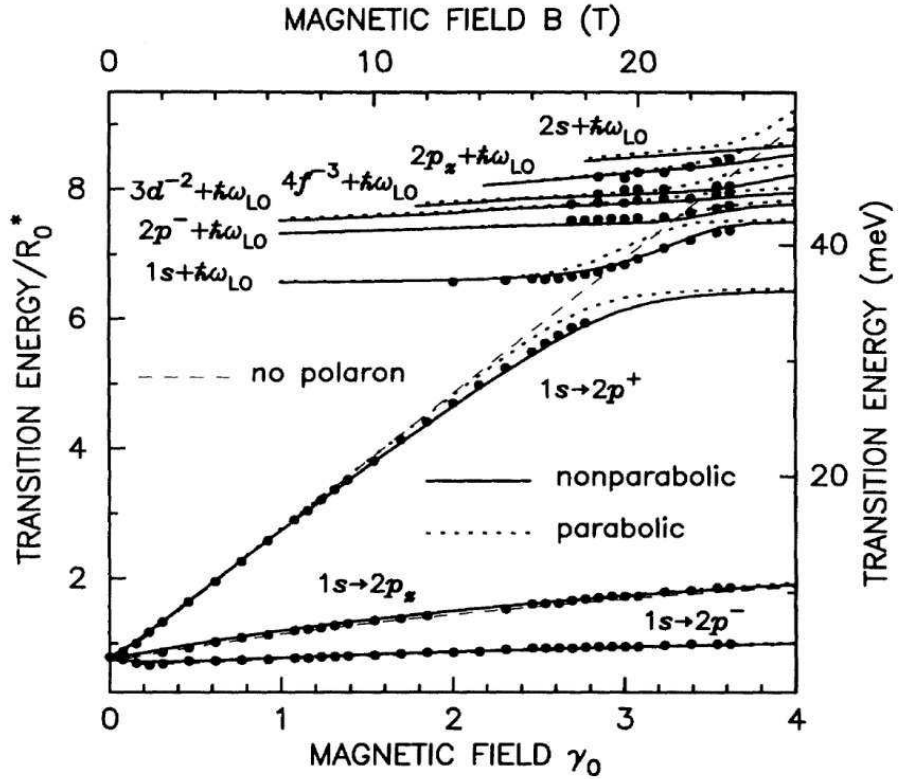}
\end{center}
\caption{The $1s \to 2p^{\pm},2p_z$ transition energies as a function of magnetic field for a donor in GaAs. {The} theoretical results {are compared to the experimental data (solid dots) \protect\cite{Cheng}} for
the following cases: (a) without the effect of polaron and band
nonparabolicity (thin dashed curves); (b) with polaron correction (dotted
curves); (c) including the effects of polaron and band nonparabolicity
(solid curves).
({Reprinted with permission from Ref. \protect\cite{SPD93}. \copyright 1993 by the American Physical Society}.)}
\label{magnetoabs}
\end{figure}

The energy spectra of polaronic systems {such} as shallow donors
(\textquotedblleft bound polarons\textquotedblright), e. g., the D$_0$ and D$%
^-$ centres, constitute the most complete and detailed polaron spectroscopy
realised in the literature, see for example Fig. \ref{magnetoabs}.

\section{Many-polaron systems in 3D and in 2D}

%\subsection{Path-integral approach to the many-polaron problem}

\subsection{Optical absorption of many-polaron systems}

\begin{figure}[b]
\begin{center}
\includegraphics[width=.8\textwidth]{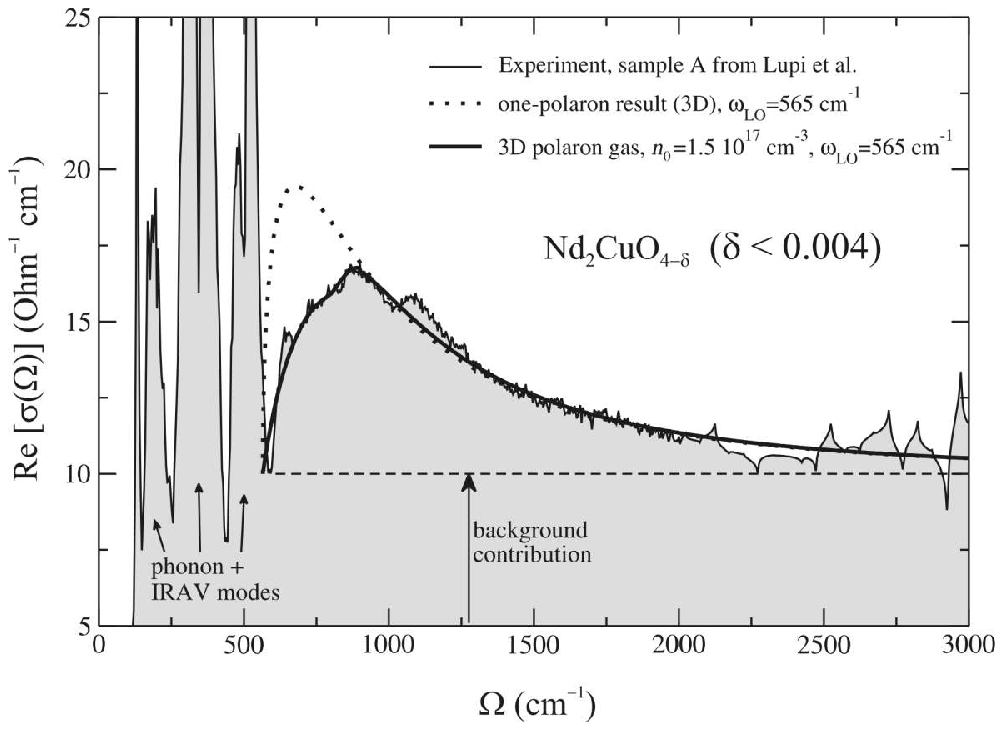}
\end{center}
\caption{The infrared optical absorption of Nd$_{2}$CuO$_{4-\protect\delta}$
($\protect\delta<0.004$) as a function of frequency. The experimental
results of Ref. \protect\cite{Lupi1999} are presented by the thin full
curve. The experimental `d-band' is clearly identified, rising in intensity
at about 600 cm$^{-1}$, peaking around 1000 cm$^{-1}$, and then decreasing
in intensity above that frequency. The dotted curve shows the single polaron
result calculated according to Ref. \protect\cite{DSG72}. The bold full
curve presents the theoretical results {of Ref. \protect\cite{TDPRB01}} for
the interacting many-polaron gas with $n_0=1.5\times10^{17}$ cm$^{-3}$, $%
\protect\alpha=2.1$ and $m=0.5m_{\mathrm{e}}$.
({Reprinted with permission from Ref. \protect\cite{TDPRB01}. \copyright 2001 by the American Physical Society}.)}
\label{fig:P24-10}
\end{figure}

\begin{figure}[htbp]
\begin{center}
\includegraphics[width=.6\textwidth]{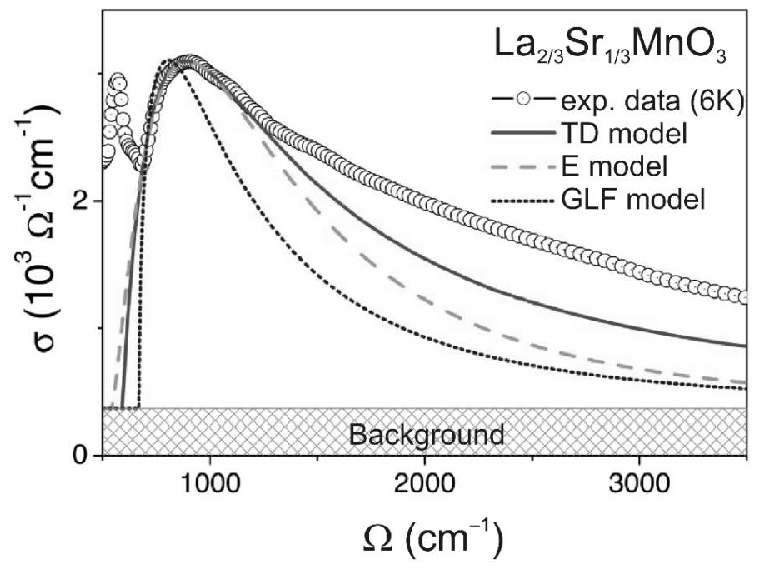}
\end{center}
\caption{Comparison of the measured mid-infrared optical conductivity in La$%
_{2/3}$Sr$_{1/3}$MnO$_{3}$ at $T=6$ K to that given by several model
calculations for $m=3m_{\mathrm{e}}$, $\protect\alpha$ of the order of 1 and
$n_0=6\times 10^{21}$ cm$^{-3}$. The one-polaron {limit} [the
weak-coupling approach by V. L. Gurevich, I. G. Lang, and Yu.~A.~Firsov {%
\protect\cite{GLF62}} (GLF model) and the phenomenological approach by D.
Emin {\protect\cite{Emin1993}} (E model)] lead to narrower polaron peaks
than a peak with maximum at \protect\linebreak $\protect\Omega \sim 900$ cm$%
^{-1}$ given by the many-polaron treatment by J. Tempere and J.~T.~Devreese
(TD model) of Ref.\,\protect\cite{TDPRB01}.
({Reprinted with permission after Ref. \protect\cite{Hartinger2004}. \copyright 2004 by the American Physical Society}.)}
\label{fig:P24-11}
\end{figure}

In Ref. \cite{TDPRB01}, starting from the many-polaron canonical
transformations and the variational many-polaron wave function introduced in
Ref.~\cite{LDB1977}, the optical absorption coefficient of a many-polaron
gas has been derived. The real part of the optical conductivity of the
many-polaron system {was} obtained in an intuitively appealing form
\begin{equation}
\mathrm{Re}[\sigma(\Omega)]=\frac{n_0}{\hbar\Omega^{3}}\frac{e^{2}}{m_{b}^{2}}%
\sum_{\mathbf{k}}k_{x}^{2}|V_{\mathbf{k}}|^{2}S(\mathbf{k},\Omega -\omega_{\mathrm{LO}}),
\label{opticabs}
\end{equation}
where $n_0$ is the density of charge carriers, $V_{\mathbf{k}}$ is the
electron-phonon interaction amplitude and $k_{x}$ is the $x$-component of
the wave vector. {Equation (\ref{opticabs}) is rigorous to order $\alpha$.} This approach to the many-polaron optical absorption allows
one to include the many-body effects in terms of the dynamical structure
factor $S(\mathbf{k},\Omega -\omega_{\mathrm{LO}})$ of the electron (or hole) system {(e.~g. in the RPA-approximation)}. The
experimental peaks in the mid-infrared optical absorption spectra of
cuprates (Fig.\thinspace\ref{fig:P24-10}) and manganites (Fig.\thinspace\ref%
{fig:P24-11}) {can be} adequately interpreted within this theory. {As seen
from Fig.\thinspace\ref{fig:P24-11}, {for the case of La$%
_{2/3}$Sr$_{1/3}$MnO$_{3}$} the many-polaron approach describes the
experimental optical conductivity better than the methods of
\cite{GLF62,Emin1993}. In Ref. \cite%
{Eagles1995}, {the interesting possibility of} coexistence of small and large polarons in the same solid {was suggested}.} %%%
%%%

\subsection{Ripplopolarons in multi-electron bubbles in liquid helium}

An interesting 2D system consists of electrons on {films of} liquid He \cite%
{SM73,9}. In this system the electrons couple to the ripplons of the liquid He, forming
``ripplopolarons''. The effective coupling can be relatively {large and} self-trapping can result.
The acoustic nature of the ripplon dispersion at long wavelengths induces the self-trapping.

Spherical shells of charged particles appear in a variety of physical systems,
such as fullerenes, metallic nanoshells, charged droplets and neutron stars. A
particularly interesting physical realization of the spherical electron gas is
found in multielectron bubbles (MEBs) in liquid helium-4. These MEBs are 0.1
$\mu$m -- 100 $\mu$m sized cavities inside liquid helium, that contain helium
vapour at vapour pressure and a nanometer-thick electron layer, anchored to the
surface of the bubble \cite{VolodinJETP26,Albrecht1987}. They exist as a result of
equilibrium between the surface tension of liquid helium and the Coulomb
repulsion of the electrons \cite{ShikinJETP27,Salomaa1981}. Recently proposed experimental
schemes to stabilize MEBs \cite{SilveraBAPS46} have stimulated theoretical
investigation of their properties (see e.~g. \cite{TemperePRL87}).

The dynamical modes of an MEB were described by considering the motion of the
helium surface (\textquotedblleft ripplons\textquotedblright) and the
vibrational modes of the electrons together. In particular, the
case when the ripplopolarons form a Wigner lattice was analyzed in Ref. \cite{TempereEPJ2003}.
The interaction energy between the ripplons and the electrons in the
multielectron bubble is derived from the following considerations: (i) the
distance between the layer electrons and the helium surface is fixed (the
electrons find themselves confined to an effectively 2D surface anchored to
the helium surface) and (ii) the electrons are subjected to a force field,
arising from the electric field of the other electrons. To study the ripplopolaron
Wigner lattice at {nonzero} temperature and for any value of the electron-ripplon coupling, the variational path-integral
approach \cite{Feynman} is used.

In their treatment of the electron Wigner lattice embedded in a polarizable
Medium, such as a semiconductor or an ionic solid, Fratini and Qu\'{e}merais
\cite{FratiniEPJB14} described the effect of the electrons on a particular
electron through a mean-field lattice potential. The (classical) lattice
potential $V_{lat}$ is obtained by approximating all the electrons acting on
one particular electron by a homogenous charge density in which a hole is
punched out; this hole is centered in the lattice point of the particular
electron under investigation and has a radius given by the lattice distance
$d$.

The Lindemann melting criterion \cite{LindemanZPhys11,Care1975} states in general that
a crystal lattice of objects (be it atoms, molecules, electrons, or
ripplopolarons) will melt when the average {displacement} of the objects {from} their
lattice site is larger than a critical fraction $\delta_{0}$ of the lattice
parameter $d$. It would be a strenuous task to calculate, from first principles,
the exact value of the critical fraction $\delta_{0}$, but for the particular
case of electrons on a helium surface, we can make use of an experimental
determination. Grimes and Adams \cite{GrimesPRL42} found that the Wigner
lattice melts when $\Gamma=137\pm15$, where $\Gamma$ is the ratio of potential
energy to the kinetic energy per electron. At temperature $T$ the average
kinetic energy {of an electron} in a lattice potential $V_{lat}$, characterized by the frequency parameter $\omega_{lat}$, is
\begin{equation}
E_{kin}={\displaystyle{\frac{\hbar\omega_{lat}}{2}}}\coth\left(
{\displaystyle{\frac{\hbar\omega_{lat}}{2k_{B}T}}}\right)  ,
\end{equation}
and the average distance that an electron moves out of the lattice site is
determined by
\begin{equation}
\left\langle \mathbf{r}^{2}\right\rangle ={\displaystyle{\frac{\hbar}%
{m_{e}\omega_{lat}}}}\coth\left(  {\displaystyle{\frac{\hbar\omega_{lat}%
}{2k_{B}T}}}\right)  ={\displaystyle{\frac{2E_{kin}}{m_{e}\omega_{lat}^{2}}}}.
\end{equation}
From this one finds that for the melting transition in Grimes and Adams'
experiment \cite{GrimesPRL42}, the critical fraction equals $\delta_{0}%
\approx0.13$. This estimate is in agreement with previous (empirical)
estimates yielding $\delta_{0}\approx0.1$ \cite{BedanovPRB49}.

Within the approach of Fratini and Qu\'{e}merais \cite{FratiniEPJB14}, the
Wigner lattice of (ripplo)\-polarons melts when at least one of the two
following Lindemann criteria are met:
\begin{equation}
\delta_{r}={\displaystyle{\frac{\sqrt{\left\langle \mathbf{R}_{cms}%
^{2}\right\rangle }}{d}}}>\delta_{0}, \label{Lind1}%
\end{equation}%
\begin{equation}
\delta_{\rho}={\displaystyle{\frac{\sqrt{\left\langle \mathbf{\rho}%
^{2}\right\rangle }}{d}}}>\delta_{0}. \label{Lind2}%
\end{equation}
where $\mathbf{\rho}$ and $\mathbf{R}_{cms}$ are, respectively, the relative
coordinate and the center of mass coordinate of the model system:
if $\mathbf{r}$ is the electron coordinate and $\mathbf{R}$ is the position
coordinate of the fictitious ripplon mass $M$, they are
\begin{equation}
\mathbf{R}_{cms}={\displaystyle{\frac{m_{e}\mathbf{r}+M\mathbf{R}}{m_{e}+M}}%
};\mathbf{\rho}=\mathbf{r}-\mathbf{R.}%
\end{equation}
The appearance of two Lindemann criteria takes into account the composite
nature of (ripplo)polarons. As follows from the physical {meaning} of the coordinates $\mathbf{\rho}$ and $\mathbf{R}_{cms}$, the first criterion
(\ref{Lind1}) is related to the melting of the ripplopolaron Wigner lattice
towards a ripplopolaron liquid, where the ripplopolarons move as a whole, the
electron together with its dimple. The second criterion (\ref{Lind2}) is
related to the dissociation of ripplopolarons: the electrons shed their dimple.

The path-integral variational formalism allows us to calculate the expectation
values $\left\langle \mathbf{R}_{cms}^{2}\right\rangle $ and $\left\langle
\mathbf{\rho}^{2}\right\rangle $ with respect to the ground state of the
variationally optimal model system.

\begin{figure}[ht]
\begin{center}
\includegraphics[width=0.5\textwidth]{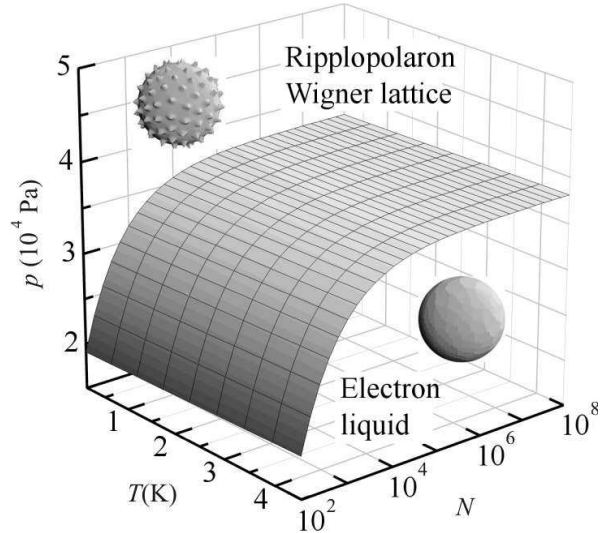}
\caption{The phase diagram for the spherical 2D layer of electrons in a MEB.
Above a critical pressure, a ripplopolaron solid (a Wigner lattice of
electrons with dimples in the helium surface underneath them) is formed. Below
the critical pressure, the ripplopolaron solid melts into an electron liquid
through dissociation of ripplopolarons.
{(Reprinted with permission from Ref. \protect\cite{TempereEPJ2003}. \copyright 2003, EDP Sciences, Societ\`{a} Italiana di Fisica, Springer.)}}
\label{PhaseD1}%
\end{center}
\end{figure}

Numerical calculation shows that for ripplopolarons in an MEB, the inequality
\begin{equation}
\left\langle \mathbf{R}_{cms}^{2}\right\rangle \ll\left\langle \mathbf{\rho
}^{2}\right\rangle
\end{equation}
is realized. {As a consequence}, the destruction of the ripplopolaron Wigner lattice in a MEB occurs
through the dissociation of ripplopolarons, since the second criterion
(\ref{Lind2}) will be fulfilled before the first (\ref{Lind1}). The results
for the melting of the ripplopolaron Wigner lattice are summarized in the
phase diagram shown in Fig. \ref{PhaseD1}.%

%EndExpansion

For {any} value of $N$, pressure $p$ and temperature $T$ in an experimentally accessible range, {Fig. \ref{PhaseD1}} shows whether the ripplopolaron Wigner lattice is present (points above the surface) or {the electron liquid} (points below the surface).
Below a critical pressure (on the order of 10$^{4}$ Pa) the ripplopolaron
solid will melt into an electron liquid. This critical pressure is nearly
independent of the number of electrons (except for the smallest bubbles) and
is weakly temperature dependent, up to the helium critical temperature 5.2 K.
This can be understood since the typical lattice potential well in which the
ripplopolaron resides has frequencies of the order of THz or larger, which
correspond to $\sim10$ K.

The new phase that was predicted {in} \cite{TempereEPJ2003}, the ripplopolaron Wigner lattice, will not be present for electrons on a flat helium surface. At the values of the pressing
field necessary to obtain a strong enough electron-ripplon coupling, the flat
helium surface is no longer stable against long-wavelength deformations
\cite{GorkovJETP18}. Multielectron bubbles, with their different ripplon
dispersion and the presence of stabilizing factors such as the energy barrier
against fissioning \cite{TemperePRB67}, allow for much larger electric fields
pressing the electrons against the helium surface. The regime of $N$, $p$, $T$
parameters suitable for the creation of a ripplopolaron Wigner lattice lies
within the regime that would be achievable in recently proposed experiments,
aimed at stabilizing multielectron bubbles \cite{SilveraBAPS46}. The
ripplopolaron Wigner lattice and its melting transition might be detected by
spectroscopic techniques \cite{GrimesPRL42,FisherPRL42} probing for example
the transverse phonon modes of the lattice \cite{DevillePRL53}.

\section{Polarons in 2D and in quasi-2D structures}

\subsection{Polarons and cyclotron resonance in quantum wells}
\label{quantum wells}

%\subsection{Polarons in the layered compound InSe}
The resonant magnetopolaron coupling \cite{JL66} in bulk manifests itself
near the LO-phonon frequency for low electron densities (see, e. g.,
Refs.\thinspace \cite{Nicholas1992,h1,brum87}) and also for higher electron
densities \cite{Swierkowski95}.

\begin{figure}[ht]
\begin{center}
\includegraphics[height=.3\textheight]{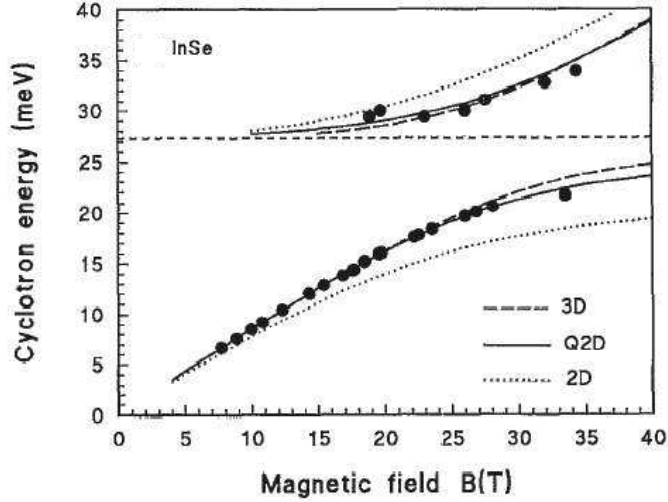}
\end{center}
\caption{The cyclotron resonance position plotted as a function of magnetic
field for InSe.
({Reprinted with permission from Ref. \protect\cite{Nicholas1992}. \copyright 1992 by the American Physical Society}.)}
\label{fig_InSe}
\end{figure}
For example, Nicholas \textit{et al.} \cite{Nicholas1992} demonstrated polaron coupling
effects using cyclotron resonance measurements in a 2DEG, which naturally
occurs in the polar semiconductor InSe. One clearly sees, over a wide range
of magnetic fields ($B=$ 18 to 34 T), two distinct magnetopolaron branches
separated by as much as 11 meV ($\sim 0.4\omega_{\mathrm{LO}}$) at
resonance (Fig.\thinspace \ref{fig_InSe}). The theoretical curves show the
results of calculations for coupling to the LO phonons in bulk (3D), sheet
(2D) and after correction for the quasi-2D character of the system {using} $%
\alpha =0.29$. The agreement
between theory and experiment is reasonable for the 3D case, but better for
the quasi-2D system, if the finite spatial extent of the 2D electron gas in
the symmetric planar layer is taken into account.

Cyclotron-resonance measurements performed on semiconductor
quantum wells with high electron density \cite{h3,Poulter2001} reveal
anticrossing near the TO-phonon frequency rather than near the LO-phonon
frequency. In Ref.\thinspace \cite{Poulter2001}, this effect is interpreted
by invoking mixing between magnetoplasmons and phonons and in terms of a
resonant coupling of the electrons with the mixed magneto-plasmon-phonon
modes.

\subsection{Cyclotron resonance in a quasi-2D many-polaron system and the role of screening}

In Refs. \cite{SPRB-2,Comment}, the CR spectra for a polaron gas in a GaAs/AlAs
quantum well are theoretically investigated, taking into account (i) the
electron-electron interaction and the screening of the electron-phonon
interaction, (ii) the magnetoplasmon-phonon mixing, (iii) the
electron-phonon interaction with all the phonon modes specific for the
quantum well under investigation. As a result of this mixing, different
magnetoplasmon-phonon modes appear in the quantum well, which give
contributions to the CR spectra.

\begin{figure}[h]
\begin{center}
\includegraphics[width=.7\textwidth]{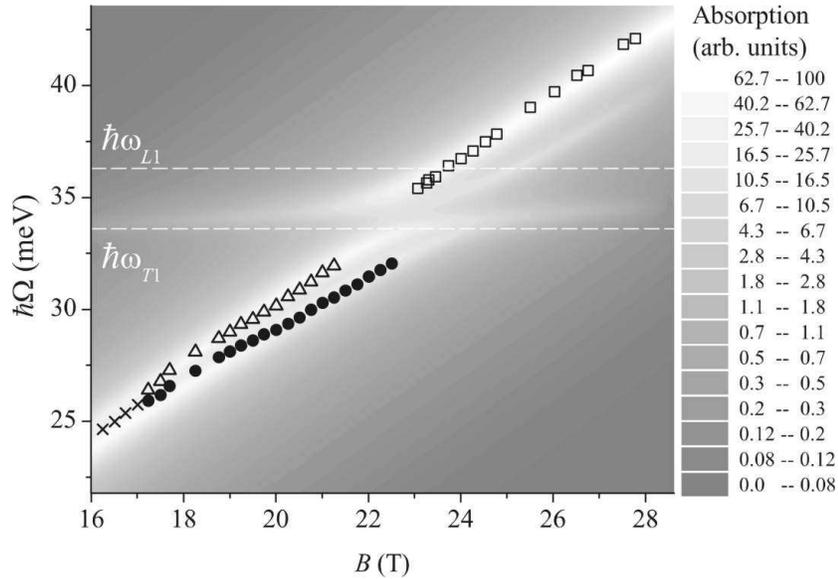}
\end{center}
\caption{Density map of the magnetoabsorption spectra for a 10-nm GaAs/AlAs
quantum well as calculated in Ref.~\protect\cite{SPRB-2}. Symbols indicate
peak positions of the experimental spectra (which are taken from Fig. 3 of
Ref. \protect\cite{Poulter2001}). Dashed lines show LO- and TO-phonon
energies in GaAs.
({Reprinted with permission after Ref. \protect\cite{SPRB-2}. \copyright 2003 by the American Physical Society}.)}
\label{fig_DensMap}
\end{figure}

It is clearly seen from Fig.\thinspace \ref{fig_DensMap}, that for a
high-density polaron gas, anticrossing of the CR spectra occurs near the
GaAs TO-phonon frequency $\omega_{T1}$ rather than near the GaAs LO-phonon
frequency $\omega_{L1}$ for both the experimental and the calculated
spectra. This effect is in contrast with the cyclotron resonance of a
low-density polaron gas in a quantum well, where anticrossing occurs near
the LO-phonon frequency. The appearance of the anticrossing frequency close
to $\omega_{T1}$ instead of $\omega_{L1}$ is due to the screening of the
electron-phonon interaction by the plasma vibrations. A similar effect
appears also for magnetophonon resonance: as shown in Ref. \cite%
{Afonin2000}, the magnetoplasmon-phonon mixing leads to a shift of the
resonant frequency of the magnetophonon resonance in quantum wells from $%
\Omega \approx \omega_{L1}$ to $\Omega \approx \omega_{T1}.$

\section{Polarons in quasi-0D structures}

\subsection{Many-polaron systems in quantum dots}
\label{quantum dots}

The ground-state energy and the optical conductivity spectra for a system
with a finite number of interacting arbitrary-coupling large polarons in a
spherical quantum dot {were} calculated using the path-integral formalism for
identical particles \cite{PRE96,PRE97,NoteKleinert}. A parabolic confinement
potential characterized by the confinement energy $\hbar \Omega _{0}$ and
with a background charge is considered. Using a generalization of the
Jensen-Feynman variational principle \cite{PRE96,NoteKleinert}, the
ground-state energy of a confined $N$-polaron system is analyzed as a
function of $N$ and of the electron-phonon coupling strength $\alpha$ \cite{dev1,SPRB}.

The total spin $S$ is analyzed as a function of the number of electrons in
the quantum dot for different values of the confinement energy, of the
coupling constant and of the ratio $\eta $ of the high-frequency and the
static dielectric constants.

Confined few-electron systems, without electron-phonon interaction, can
exist in one of {\it two} phases: a spin-polarized state and a state obeying
Hund's rule, depending on the confinement frequency (see, e.~g.,
Ref.\thinspace \cite{koskinen}). For interacting few-polaron systems with $%
\alpha \geq 3$, besides the above two phases, a {\it third}
phase {may occur}: the state with minimal spin.

In Ref. \cite{SPRB} the memory-function approach has been extended to a
system of arbitrary-coupling interacting polarons, confined to a parabolic
confinement potential. The applicability of the parabolic approximation is confirmed by the fact that the self-induced polaronic potential, created by the polarization cloud around
an electron, is rather well described by a parabolic potential whose
parameters are determined by a variational method. For weak coupling, our
variational method is at least of the same accuracy as the perturbation
theory, which results from our approach at a special choice of the variational
parameters. For strong coupling, the interplay of the electron-phonon
interaction and the Coulomb correlations within a confinement potential can
lead to the {clustering} of polarons in multi-polaron systems. {Our} choice of the
model variational system is {justified} because of this trend, {which occurs} in a many-polaron system with arbitrary $N$, for a finite confinement strength.

\begin{figure}[h]
\begin{center}
\includegraphics[width=0.7\textwidth]
{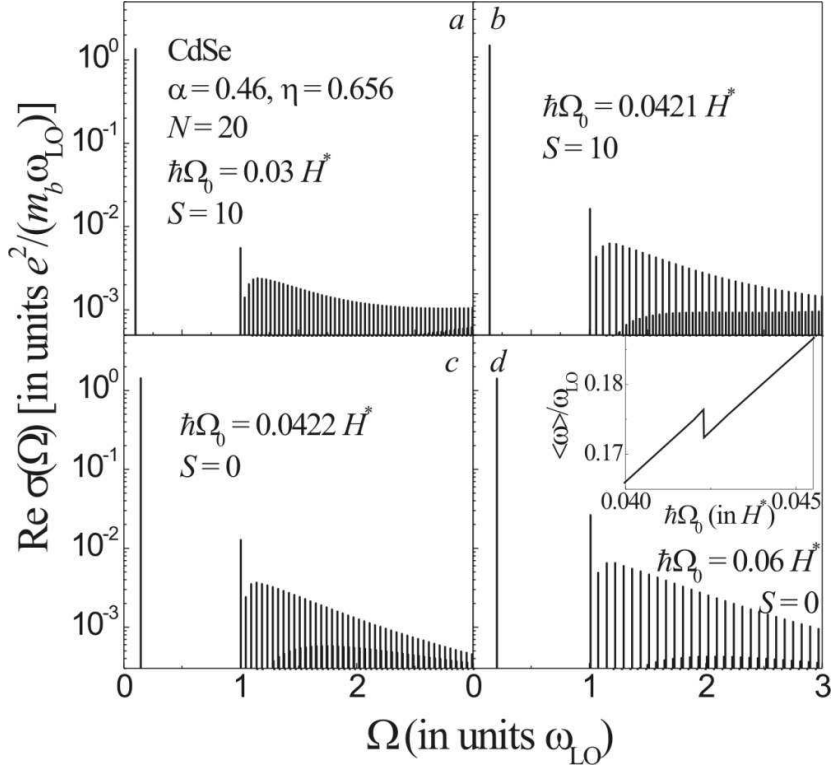}%
\caption{Optical conductivity spectra of $N=20$ interacting polarons in CdSe
quantum dots with $\alpha=0.46$, $\eta=0.656$ for different confinement
energies close to the transition from a spin-polarized ground state to a
ground state obeying Hund's rule. \emph{Inset}: the first
frequency moment $\left\langle \Omega\right\rangle $ of the optical
conductivity as a function of the confinement energy.
({Reprinted with permission from Ref. \protect\cite{SPRB}. \copyright 2004 by the American Physical Society}.)}
\label{Spectra}%
\end{center}
\end{figure}
%EndExpansion

The shell filling schemes {of} an $N$-polaron system in a quantum dot can
manifest themselves in the optical conductivity. In
Fig.\thinspace\ref{Spectra}, optical conductivity spectra for $N=20$ polarons
are presented for a quantum dot with the parameters of CdSe: $\alpha=0.46,$
$\eta=0.656$ \cite{Kartheuser1972} and with different values of the
confinement energy $\hbar\Omega_{0}$.\footnote[7]{For the numerical
calculations, we use effective atomic units, where $\hbar,$ the electron band
mass $m_{b}$ and $e/\sqrt{\varepsilon_{\infty}}$ have the numerical value of
1. This means that the unit of length is the effective Bohr radius
$a_{B}^{\ast}=\hbar^{2}\varepsilon_{\infty}/\left(  m_{b}e^{2}\right)  $,
while the unit of energy is the effective Hartree $H^{\ast}=m_{b}e^{4}/\left(
\hbar^{2}\varepsilon_{\infty}^{2}\right)  $.} In this case, the spin-polarized
ground state changes to the ground state satisfying Hund's rule with
increasing $\hbar\Omega_{0}$ in the interval $0.0421H^{\ast}<\hbar\Omega
_{0}<0.0422H^{\ast}$.%

In the inset to Fig.\thinspace\ref{Spectra}, the first frequency moment of the
optical conductivity
\begin{equation}
\left\langle \Omega\right\rangle \equiv\frac{\int_{0}^{\infty}\Omega
\mathrm{Re}\sigma\left(  \Omega\right)  d\Omega}{\int_{0}^{\infty
}\mathrm{Re}\sigma\left(  \Omega\right)  d\Omega}, \label{Moment}%
\end{equation}
as a function of $\hbar\Omega_{0}$ shows a \emph{discontinuity}, at the value
of the confinement energy corresponding to the change of the shell filling
schemes from the spin-polarized ground state to the ground state obeying
Hund's rule. This discontinuity {could} be observable in optical measurements.

\begin{figure}
[ptbh]
\begin{center}
\includegraphics[width=0.5\textwidth]
{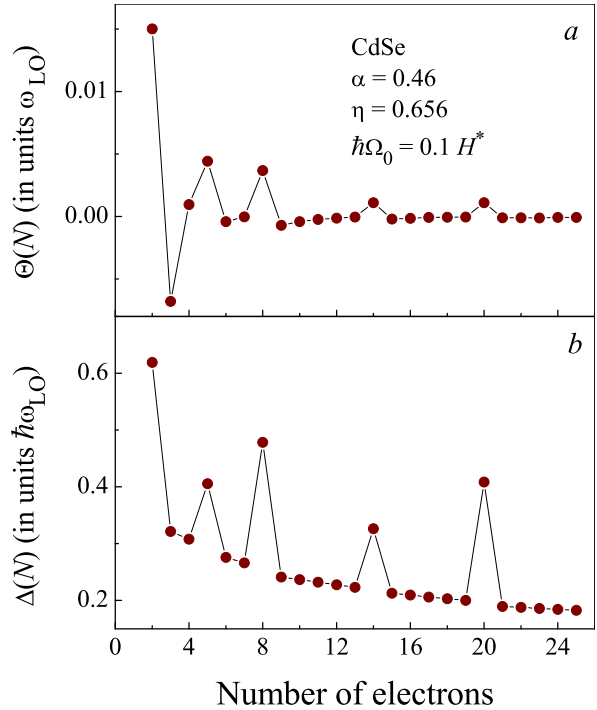}%
\caption{The function $\Theta\left(  N\right)  $ and the addition energy
$\Delta\left(  N\right)  $ for systems of interacting polarons in CdSe quantum
dots with $\alpha=0.46$, $\eta=0.656$ for $\hbar\Omega_{0}=0.1H^{\ast}$.
({Reprinted with permission from Ref. \protect\cite{SPRB}. \copyright 2004 by the American Physical Society}.)}
\label{Moments}%
\end{center}
\end{figure}

The shell structure for a system of interacting polarons in a quantum dot is
clearly revealed when analyzing the addition energy and the first frequency
moment of the optical conductivity in parallel. In Fig \ref{Moments}, we show
both the function
\begin{equation}
\Theta\left(  N\right)  \equiv\left.  \left\langle \Omega\right\rangle
\right\vert _{N+1}-2\left.  \left\langle \Omega\right\rangle \right\vert
_{N}+\left.  \left\langle \Omega\right\rangle \right\vert _{N-1},
\label{Theta}%
\end{equation}
and the addition energy%
\begin{equation}
\Delta\left(  N\right)  =E^{0}\left(  N+1\right)  -2E^{0}\left(  N\right)
+E^{0}\left(  N-1\right)  . \label{Add}%
\end{equation}
for interacting polarons in a 3D CdSe quantum dot.%

%EndExpansion

As seen from Fig \ref{Moments}, distinct peaks appear in $\Theta\left(
N\right)  $ and $\Delta\left(  N\right)  $ at the \textquotedblleft magic
numbers\textquotedblright\ corresponding to closed-shell configurations at
$N=8,20$ and to half-filled-shell configurations at $N=5,14$. We see that each
of the peaks of $\Theta\left(  N\right)  $ corresponds to a peak of the
addition energy. {The peak positions for the addition energy of interacting polarons in a 2D parabolic quantum dot \cite{SPRB} agree well with the experimental results for the addition energies of cylindrical GaAs quantum dots \cite{Tarucha}.}

\subsection{Non-adiabaticity of polaronic excitons in semiconductor quantum dots.
Photoluminescence and Raman scattering of polarons in quantum dots}
\label{optical response in quantum dots}

Interest in the optical properties of quantum dots has been continuously
growing {i.~a.} because of the prospects of these structures for
optoelectronic applications. Recent measurements~\cite{2,3,4,5} of
photoluminescence of self-assembled InAs/GaAs quantum dots, reveal {unexpectedly} high
probabilities of the phonon-assisted transitions. Attempts to interpret some
of these experiments on the basis of the adiabatic theory meet with
difficulties. For spherical CdSe quantum dots, e. g., the Huang-Rhys factor $%
S$ calculated within the adiabatic approximation takes values which are
significantly (by one or two orders of magnitude) smaller than the so-called
``experimental Huang-Rhys factor''. {This ``experimental Huang-Rhys factor'' is} determined from the ratio of the measured
intensities of the phonon satellites. In the framework of the adiabatic
approach, various mechanisms, which cause the separation of the electron and
hole charges in space~\cite{3,6,7}, are commonly considered as a possible
origin for the unexpectedly large Huang-Rhys factor.

It has been shown {in Ref.} \cite{PRB57-2415}, that non-adiabaticity of exciton-phonon systems
in some quantum dots drastically enhances the efficiency of the exciton-phonon interaction,
especially when the exciton levels are separated with energies close to the
phonon energies. Also ``intrinsic'' excitonic degeneracy can lead to
enhanced efficiency of the exciton-phonon interaction. The effects of
non-adiabaticity are important to interpret the surprisingly high
intensities of the phonon `sidebands' observed in the optical absorption,
the photoluminescence and the Raman spectra of {some} quantum dots. {Major} deviations of the {observed} oscillator strengths of {some} phonon-peak sidebands from the Franck-Condon progression, which is prescribed by the commonly used adiabatic approximation, find a natural explanation within {our} non-adiabatic approach, {introduced in \cite{PRB57-2415}, see} \cite%
{nanot,raman2002,PSS03,Fonoberov2004}.

In Ref. \cite{PRB57-2415}, a method was proposed to calculate the optical absorption
spectrum for a spherical quantum dot taking into account the
non-adiabaticity of the exciton-phonon system. This approach has been
further refined in Ref.\thinspace \cite{STACK}: for the matrix elements of
the evolution operator a closed set of equations has been obtained using a
diagrammatic technique. This set describes the effect of non-adiabaticity
both on the intensities and on the positions of the absorption peaks.

\begin{figure}[tbp]
\begin{center}
\includegraphics[width=.45\textwidth]{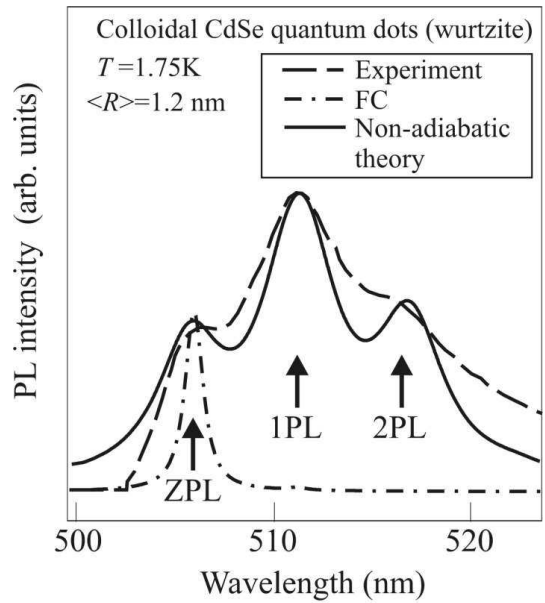}
\end{center}
\caption{Photoluminescence spectra of colloidal spherical CdSe quantum dots with wurtzite structure.
The dashed curve represents the experimental data~from Ref. \cite{Nirmal}. The dash-dotted curve
displays a result of the adiabatic approximation -- a Franck-Condon progression with Huang-Rhys factor
$S=0.06$ as calculated by S. Nomura and S. Kobayashi \cite{6}. The solid curve results from the non-adiabatic theory.
({Reprinted with permission after Ref. \protect\cite{PRB57-2415}. \copyright 1998 by the American Physical Society}.)}
\label{fig:P24-12}
\end{figure}

\begin{figure}[h]
\begin{center}
\includegraphics[width=1.0\textwidth]{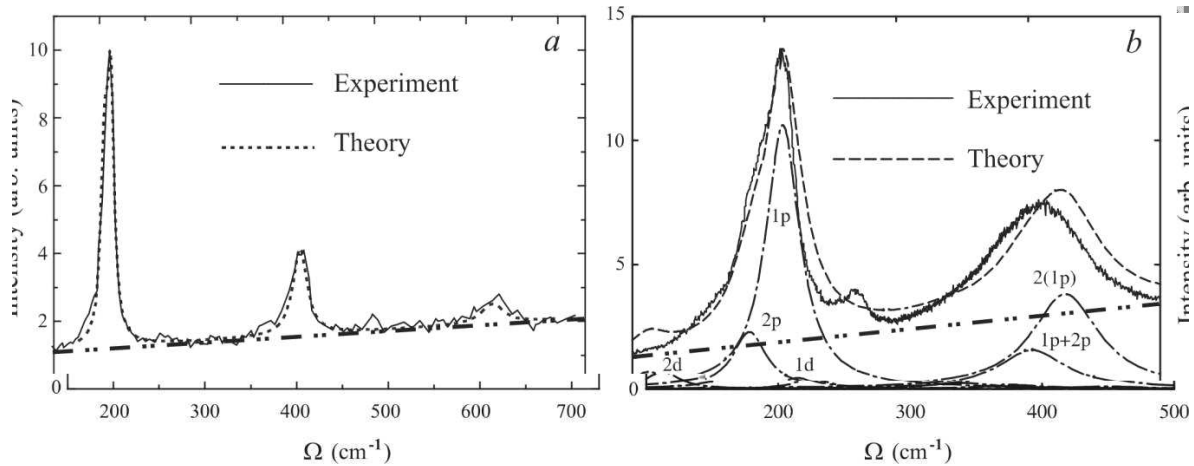}
\end{center}
\caption{Resonant Raman scattering spectra of an ensemble of CdSe quantum dots with average radius 2 nm at $T=77$ K (panel $a$)
and of PbS quantum dots with average radius 1.5 nm at $T=4.2$ K (panel $b$). The dash-dot-dot curves show the luminescence background.
The dash-dot curves in panel $b$ indicate contributions of phonon modes (classified in analogy with electron states in a hydrogen atom)
into the Raman spectrum. The dashed curves represent experimental Raman spectra.
({Reprinted with permission after Ref. \protect\cite{raman2002}. \copyright 2002 by the American Physical Society}.)}
\label{fig:P24-13}
\end{figure}

While the adiabatic approximation {predicts} negligibly low intensities of one- and two-phonon sidebands,
{our} non-adiabatic theory allows for a quantitative interpretation of the observed high intensity of
the phonon sidebands in the photoluminescence \index{photoluminescence spectra of quantum dots}
(Fig.~\ref{fig:P24-12}) and Raman \index{Raman spectra of quantum dots} (Fig.~\ref{fig:P24-13})
spectra of some quantum dots, in agreement with experiment.

From the comparison of the spectra obtained in the adiabatic approximation
with those resulting from the non-adiabatic approach, the following effects
of non-adiabaticity are revealed. First, the \textit{polaron shift} of the
zero-phonon lines with respect to the bare-exciton levels is larger in the
non-adiabatic approach than in the adiabatic approximation. Second, there is
a strong \textit{increase of the intensities of the phonon satellites} as
compared to those {derived} {within} the adiabatic approximation. Third, in the
optical absorption spectra found within the non-adiabatic approach, phonon satellites {appear, that are} related to \textit{non-active bare exciton states}.
Fourth, the optical-absorption spectra demonstrate the crucial role of
\textit{non-adiabatic mixing} of different exciton and phonon states in
quantum dots. This results in a rich structure of the absorption spectrum of
the exciton-phonon system \cite{GBFD2001,nanot,PSS03}. Similar conclusions
about the pronounced influence of the exciton-phonon interaction on the
optical spectra of quantum dots have been {subsequently} formulated in Ref.~\cite{VFB2002} in terms of a strong coupling regime for excitons and LO phonons.
Such a strong coupling regime is a particular case of the non-adiabatic mixing, {studied by us \cite{PRB57-2415,GBFD2001},} related to a (quasi-) resonance which arises when the spacing between
exciton levels is close to the LO phonon energy. The large enhancement of
the two-phonon sidebands in the luminescence spectra as compared to those
predicted by the Huang-Rhys formula, which was explained in Ref. \cite%
{PRB57-2415,GBFD2001} by non-adiabaticity of the exciton-phonon system in certain
quantum dots, has been reformulated in Ref. \cite{VFB2002} in terms of the Fr%
\"{o}hlich coupling between product states with different electron and/or
hole states

Due to non-adiabaticity, multiple absorption peaks appear in spectral ranges
characteristic for phonon satellites. From the states, which correspond to
these peaks, the system can rapidly relax to the lowest emitting state.
Therefore, in the photoluminescence excitation (PLE) spectra of {specific} quantum dots, pronounced peaks can be expected in spectral ranges characteristic for
phonon satellites. New experimental evidence of the enhanced phonon-assisted
absorption due to effects of non-adiabaticity has been provided by
PLE measurements on single self-assembled InAs/GaAs {quantum dots} \cite{lem01} and InGaAs/GaAs quantum dots \cite{zre01}.

\section*{Acknowledgements}
I like to thank V. M. Fomin for discussions during the
preparation of this manuscript. I acknowledge also discussions with S. N.
Klimin, V. N. Gladilin, A. S. Mishchenko, V. Cataudella, G. De Filippis, R.
Evrard, F. Brosens, {L. Lemmens} and J. Tempere.
This work has been supported by the GOA
BOF UA 2000, IUAP, FWO-V projects G.0306.00, G.0274.01N, G.0435.03, the WOG
WO.035.04N (Belgium) and the European Commission SANDiE Network of
Excellence, contract No. NMP4-CT-2004-500101.

\end{document}